\documentclass[amsmath,superscriptaddress,showpacs,prl,twocolumn] {revtex4}
\usepackage{bm}
\usepackage{enumerate}
\usepackage[dvips]{graphicx}
\usepackage[dvips]{epsfig}

\begin{document}

\makeatletter
\newcommand{\Rmnum}[1]{\expandafter\@slowromancap\romannumeral #1@}
\makeatletter

\title{Fractional Quantum Hall Effect from Phenomenological Bosonization}
\author{Vladimir A. Zyuzin}
\affiliation{Department of Physics, The University of Texas at Austin, Austin, TX 78712, USA}
\begin{abstract}
In this paper we propose a model of the fractional quantum Hall effect within conventional one-dimensional bosonization. It is shown that in this formalism the resulting bosonized fermion operator corresponding to momenta of Landau gauge wave function is effectively two-dimensional. At special filling factors the bulk gets gapped, and the theory is described by a sine-Gordon model. The edges are shown to be gapless, chiral, and carrying a fractional charge. The hierarchy of obtained fractional charges is consistent with existing experiments and theories. It is also possible to draw a connection to composite fermion description and to the Laughlin many-body wave function.
\end{abstract}
\maketitle

{\it Introduction. -}
The fractional quantum Hall effect (FQHE) is the most striking effect in condensed matter physics. Since the discovery of the FQHE \cite{Tsui} there were a number of theories explaining the fractional electronic states. Laughlin \cite{Laughlin83} constructed variational many-body wave-function and showed that it describes the ground state of interacting two-dimensional electron gas in a quantizing magnetic field. The ground state is an incompressible fermionic liquid with fractional excitations. The Chern-Simons effective field theory was then shown to describe the fractional quantum Hall state \cite{Girvin_MacDonald, Kalmeyer}. Also, the conformal field theory approach was developed to describe many-body wave-function \cite{MooreGreen}. Hierarchy of fractional numbers that correspond to the Hall plateaux was developed in the series of works \cite{Haldane83,Halperin84,Jain89,Read90}. In another work \cite{Kane,Kane2}, an attempt to understand the FQHE was made by using sliding Luttinger liquid. Present paper generalizes \cite{Kane,Kane2} for the case of two-dimensional electron gas in quantizing magnetic field.

{\it A two-dimensional electron gas in a magnetic field. -} Let us study a two-dimensional electron gas in a perpendicular magnetic field and choose Landau gauge for vector potential, $A_{x}=-By,~A_{y}=A_{z}=0$, where $B$ is a magnetic field. The Hamiltonian is 
\begin{equation}\label{Landau}
H=\frac{1}{2m_{e}}\left({\hat p}_{x} + \frac{eB}{c} y\right)^{2} + \frac{{\hat p}_{y}^{2}}{2m_{e}},
\end{equation}
eigenfunctions corresponding to this Hamiltonian are Landau wave-functions:
\begin{equation}\label{Landauwf}
\Psi_{n,y_{0}} = e^{ip_{x}x/\hbar}\chi_{n}(y-y_{0}),
\end{equation}
where $\chi_{n}(y) = \frac{1}{\pi^{1/4}a_{B}^{1/2}\sqrt{2^{n}n!}}\exp\left(-\frac{y^{2}}{2a_{B}^{2}}\right)H_{n}\left(\frac{y}{a_{B}}\right)
$. The eigenfunctions (\ref{Landauwf}) correspond to the energy values $E_{n}=\hbar\omega_{B}(n+\frac{1}{2})
$, where $a_{B}=\sqrt{\frac{\hbar}{m_{e}\omega_{B}}}$, $\omega_{B}=\frac{|e|B}{m_{e}c}$, $y_{0}=\frac{cp_{x}}{|e|B}$ is a position, $H_{n}$ are Hermite polynomials, $m_{e}$ is an electron mass, $e$ is charge, and $c$ is a speed of light. The energy levels take discrete values and are degenerate \cite{LL3}. Throughout the paper, we will be working with zeroth Landau level ($n=0$), set $|e|=e$, and assume spinless electrons.

Let us assume that the positions $y_{j}$ are discrete, and let us set the distance between two neighboring positions as $d_{y}$. Throughout the paper, we are going to label momenta as $p_{j}$ and mean that they correspond to the positions $y_{j}=jd_{y}$, where $j$ is an integer. When the Coulomb interaction
\begin{equation}\label{Coulomb}
H_{C} = \frac{1}{2}\int d{\bf r}d{\bf r}^{\prime}~ V({\bf r}-{\bf r}^{\prime}) \rho({\bf r})\rho(\bf r^{\prime}),
\end{equation}
where $\rho({\bf r})$ is the electron density, is present in the system, wave-functions (\ref{Landauwf}) are no longer the eigen-states of the total Hamiltonian. Instead of solving for the new eigenfunctions of interacting Hamiltonian (\ref{Landau}) and (\ref{Coulomb}), we are going to construct a trial wave-function based on the following two analogies.

It is known that the tunneling current between two perfect two-dimensional gases is non zero due to the Coulomb interactions \cite{MacDonald, Zyuzin}. Coulomb interaction lifts the orthogonality between the electron states in two gases, allowing for them to mix. And we notice that in our problem, the non-interacting electron states with different momentum, which are defined by (\ref{Landauwf}), are orthogonal to each other. Therefore, one should expect that, due to the Coulomb interaction, there will be a mixing of states with different momentum.

We then notice that a wave-function (\ref{Landauwf}) with a given momentum $p_{j}$ has a sharp peak at $y_{j}$ . Therefore, an electron corresponding to this state can be seen as one-dimensional. That allows one to use phenomenological one-dimensional bosonization, motivated by the knowledge that the effect of electron interactions can be treated non-perturbatively within this approach \cite{Haldane81, Cazalilla, Giamarchi}. Let us show that the bosonization in quantizing magnetic field is effectively two-dimensional.

{\it Phenomenological bosonization. -} In the approach of phenomenological one-dimensional bosonization, the electron density is described by the harmonics of density fluctuations
\begin{equation}\label{density}
\rho(x) = \left( \rho_{0} - \frac{1}{\pi}\nabla\phi(x) \right)\sum_{n=-\infty}^{+\infty} e^{i2n(\pi \rho_{0}x - \phi(x))},
\end{equation}
where $\phi(x)$ is a smooth function describing the deviation from the homogeneous density distribution given by $\rho_{0}$. Following the lines of \cite{Haldane81, Cazalilla, Giamarchi}, we can present a fermion operator as
\begin{eqnarray}\label{fermionwf}
&\Psi^{\dag}(x) = \left(\rho_{0} - \frac{1}{\pi}\nabla \phi(x)  \right)^{1/2} \\
&\times\sum_{n=-\infty}^{+\infty} e^{i(2n+1)(\pi\rho_{0}x - \phi(x))} e^{ -i\theta(x)}, \nonumber
\end{eqnarray}
where $\theta(x)$ is a conjugate to $\phi(x)$ bosonic field. This fermion operator describes strictly one-dimensional fermions.

\begin{figure}
\includegraphics[width=0.9\linewidth]{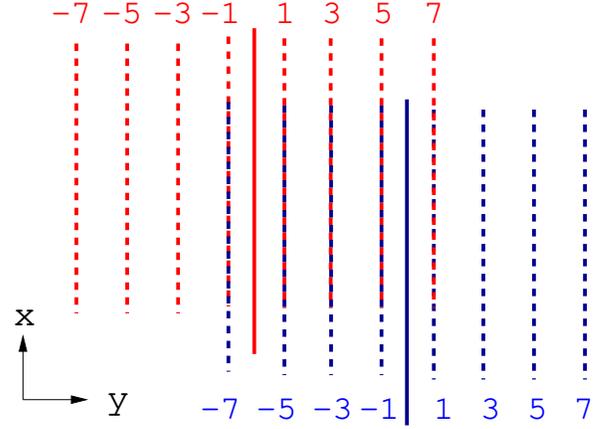}
\caption{(color online) Solid lines represent positions of two Landau wave-function (\ref{Landauwf}) with two different momentum, red corresponds  $p_{j}$, while blue - to $p_{j+1}$. Dashed lines with odd integers represent positions corresponding to the $(2n+1)$ harmonic of the density fluctuations of (\ref{bosonizedwf}). For example, $"3"$ stands for $n=-2$ part of (\ref{bosonizedwf}). Again, red dashed line corresponds to the momentum $p_{j}$, while blue to $p_{j+1}$. Any bosonized fermion operator covers the entire plane with its harmonics. This figure describes $\nu=1/3$ filling factor, and the Laughlin state is shown: match of the (3,-3) harmonics. In this notation $"3"$ stands for a harmonic $n=-2$ of $p_{j}$, and $"-3"$ for $n=1$ of $j+1$. \label{fig1}}
\end{figure}

Let us now adopt this phenomenological bosonization to construct a fermion operator for two-dimensional electrons in a quantizing magnetic field. We have to bosonize electrons with a fixed momentum $\exp(ip_{j}x/\hbar)$. Fixing the momentum dictates choosing a fermion operator as
\begin{equation}\label{firststep}
\Psi_{b,j}^{\dag} \propto e^{-ieBy_{j}x/c\hbar}.
\end{equation}
The bosonized fermion operator should have a similar form to the expression (\ref{fermionwf}). If we just multiply (\ref{fermionwf}) with (\ref{firststep}), the harmonics will shift the phase $(-ieBy_{j}/c\hbar)$ by a factor of $i(2n+1)\pi\rho_{0}$. Therefore, we write the bosonized fermion operator as
\begin{eqnarray}\label{bosonizedwf}
&\Psi_{b,j}^{\dag}(x,y) = \eta_{j} e^{-ieBy_{j}x/c\hbar}\left(\rho_{0} - \frac{1}{\pi}\nabla \phi_{j}(x)  \right)^{1/2} \\
&\times \sum_{n=-\infty}^{+\infty} \chi_{0}(y-{\tilde y}_{j,n})e^{i(2n+1)(\pi\rho_{0}x - \phi_{j}(x)) } e^{-i\theta_{j}(x)}, \nonumber
\end{eqnarray}
where $\eta_{j}$ is a Klein factor which assures the fermionic anti-commutation relationship between the states with different momentum, we define them as $\{\eta_{j},{\eta_{j^{\prime}}}\} = 2\delta_{jj^{\prime}}$, and $\eta_{j}\eta_{j^{\prime}} = i$. The
\begin{equation}\label{guiding}
{\tilde y}_{j,n} = y_{j} - (2n+1)\nu d_{y}/2
\end{equation}
is a position of electron state which contributes the $(2n+1)$ harmonic to the fermion operator with a fixed momentum $p_{j}$ (see Fig. \ref{fig1}). Here $\nu = \frac{\rho_{2d}hc }{eB}$ is a filling factor of a two-dimensional electron gas in a magnetic field, where $h=2\pi\hbar$, $\rho_{0}$ is an effective one-dimensional electron density and $\rho_{2d}=\rho_{0}/d_{y}$ is a two-dimensional electron density.

Being borrowed from one-dimensional electron picture, this fermion operator describes electron density fluctuations of a state with a fixed momentum at different $y$ coordinates. This makes the (\ref{bosonizedwf}) effectively two-dimensional. In it's description the higher harmonics of the density fluctuations are analogous to the composite fermions \cite{Jain89} with guiding centers defined by (\ref{guiding}). Here, the flux attachment procedure is expressed through a shift of the position. A possible justification for the use of bosonization approach to a problem of FQHE is the existence of one-dimensional presentation of Laughlin many-body wave function \cite{Azuma, Panigrahi}. And we speculate that a many-body Laughlin wave function can be mathematically transformed to (\ref{bosonizedwf}).

Let us now using (\ref{bosonizedwf}) construct a Hamiltonian describing density fluctuations of the system.

{\it Hamiltonian of density fluctuations. -} Let us first construct a Hamiltonian describing fluctuations of the electron density of a state with a fixed $j$. The diagonal matrix element of Hamiltonian (\ref{Landau}) on the fermion operator ({\ref{bosonizedwf}}) with a given $j$, taken with $(n=0,~-1)$, is
\begin{equation}\label{kin}
H^{kin}_{j}=\frac{\hbar^2 \rho_{0}}{m_{e}} \int dx~\left[ (\nabla \phi_{j}(x))^{2} + (\nabla \theta_{j}(x))^{2} \right].
\end{equation}
Another part of the Hamiltonian is due to the Coulomb interaction (\ref{Coulomb}) between electrons with $j$. Forward scattering between electrons is obtained if one plugs (\ref{bosonizedwf}) with $n=0$ into the expression of Coulomb interaction (\ref{Coulomb})
\begin{equation}\label{fs}
H^{fs}_{j} = \frac{g_{1}}{2\pi} \int dx~(\nabla \phi_{j}(x) )^{2},
\end{equation}
where $g_{1} \cong 2c_{1} e^{2}/\pi\epsilon_{0}$, $\epsilon_{0}$ is the dielectric constant, and $c_{1}$ is the constant which depends on the large distance cut-off. The total Hamiltonian of electron density fluctuations then takes a form
\begin{equation}\label{kinfs}
H_{j} = \frac{1}{2\pi} \int dx~\left[v(\nabla\theta_{j}(x))^{2} + (v+g_{1})(\nabla\phi_{j}(x) )^{2} \right],
\end{equation}
here $v=2\pi\hbar^2 \rho_{0}/m_{e}$. Now let us include the Coulomb interaction between electron states with different $j$. The forward-scattering reads
\begin{equation}\label{ifs}
H^{fs}_{j,m} = \frac{g_{2}}{2\pi}\int dx~ (\nabla\phi_{j}(x))(\nabla\phi_{j+m}(x)),
\end{equation}
where $m$ is a non-zero integer denoting the number of a neighbor. Backscattering between the states with different $j$ momentum gives
\begin{equation}\label{ibs}
H^{bs}_{j,m}=\frac{2\rho_{0}^{2}}{(2\pi)^2}g_{3}\int dx~\cos\left(2\phi_{j}(x) - 2\phi_{j+m}(x)\right).
\end{equation}
To estimate $g_{2}$ and $g_{3}$ one has to calculate integral of (\ref{Coulomb}) taken with $x, ~y$ parts of (\ref{bosonizedwf}). The latter constant exponentially decays with $m$. To complete the description of the Hamiltonian of density fluctuations, we need to include non-diagonal matrix elements of the Hamiltonian (\ref{Landau}).

{\it Fractional states. -} The non-diagonal matrix elements are obtained by acting with $\frac{{\hat p}^{2}_{y}}{2m_{e}}$ of the Hamiltonian (\ref{Landau}) on the (\ref{bosonizedwf}) of a momentum $p_{j+m}$, and taking the overlap with the neighboring $p_{j}$ momentum
\begin{equation}\label{nondiagonal}
\frac{-\hbar^{2}}{2m_{e}}\int dxdy~\Psi_{b,j}^{\dag}(x,y)\partial_{y}^{2}\Psi_{b,j+m}(x,y).
\end{equation}
Let us show how fractional charges appear at $\nu=1/3$ filling. In this case, we need to set $m=1$ and pick the harmonics of (\ref{bosonizedwf}) in the expression above as $(3,~-3)$ (see Fig. \ref{fig1} for explanation of the latter notation)
\begin{eqnarray}\label{laughlin}
& t_{1}\rho_{0}\int dx~(\Psi_{b,j}^{(n=-2)})^{\dag} \Psi_{b,j+1}^{(n=1)}+h.c. = t_{1}\rho_{0}\eta_{j}\eta_{j+1} \\
&\times \int dx~ e^{-i(6\pi\rho_{0}-\frac{eB}{c\hbar}d_{y})x }e^{i3(\phi_{j} + \phi_{j+1})}e^{i(\theta_{j+1}-\theta_{j})} + h.c. \nonumber
\end{eqnarray}
The constant $t_{1}$ is defined as $t_{1} = \frac{-\hbar^{2}}{2m_{e}}\int dy~\chi_{0}(y-{\tilde y}_{j,-2})\partial_{y}^{2}\chi_{0}(y-{\tilde y}_{j+1,1})$. Another, equal, contribution to this matrix element will come from the $\propto (y-{\tilde y}_{j,1})^{2}$ term of the Hamiltonian (\ref{Landau}), and we are going to double the $t_{1}$ due to that. For this term to be relevant, the oscillating part $\exp\left[-i\left(6\pi\rho_{0} - \frac{eB}{c\hbar}d_{y} \right)x\right]$ has to be zero. This sets a condition on the filling factor
\begin{equation}\label{filling}
\nu = \frac{\rho_{2d}hc }{eB}= \frac{1}{3}.
\end{equation}
Notice that the $\nu=1/3$ makes the $y-$coordinates of selected in (\ref{laughlin}) harmonics, $(3,~-3)$, to match (see Fig. \ref{fig1}). The value of the overlap in this case is $t_{1} =\hbar\omega_{B}/2$. Let us rewrite the expression (\ref{laughlin}) in the form
\begin{equation}\label{laughlin2}
H_{j,L} = \hbar\omega_{B}\rho_{0}\int dx~ \cos(3(\phi_{j} + \phi_{j+1}))\sin(\theta_{j}-\theta_{j-1}).
\end{equation}
One can show that all matchings of harmonics to the left, such as $(-1,~-7)$ and $(1,~-5)$ on Fig. \ref{fig1}, vanish because the oscillating part in the overlap between them can not be put to zero. However, the matchings of the harmonics to the right, such as $(5,~-1)$ and $(7,~1)$, do not have an oscillating part. We are going to show that they vanish, and the reason will be clear from the following analysis. Let us study the following Hamiltonian, which is a sum of (\ref{kinfs}), (\ref{ifs}), (\ref{ibs}) and (\ref{laughlin2}) taken with $m=1$
\begin{equation}\label{totalham}
{\cal H} = \sum_{j} \left[ H_{j}+H^{fs}_{j,1}+H^{bs}_{j,1} + H_{j,L} \right].
\end{equation}
In this case the Coulomb interaction constants are $g_{2} \approx g_{1}$ and $g_{3} \approx 0.006\frac{e^{2}}{\epsilon_{0}}$.
It can be shown that the Hamiltonian (\ref{totalham}) decomposes into a sum of coupled symmetric and anti-symmetric sectors of neighboring momentum, with corresponding bosonic fields $\phi_{j+1/2, \pm} = (\phi_{j} \pm \phi_{j+1})/\sqrt{2}$ and $\theta_{j+1/2,\pm} = (\theta_{j} \pm \theta_{j+1})/\sqrt{2}$. We can then describe the whole system by focusing on just one pair of neighboring momentum (below we omit $j+1/2$ index for simplicity), which is described by $H_{pair} = H_{+} + H_{-} + H_{L}$, where
\begin{eqnarray}\label{two}
&H_{+} = \frac{1}{2\pi}\int dx~\left[u_{+}K_{+}(\nabla\theta_{+})^{2}  + \frac{u_{+}}{K_{+}}(\nabla\phi_{+})^{2} \right], \nonumber\\
&H_{-} = \frac{1}{2\pi}\int dx~\left[u_{-}K_{-}(\nabla\theta_{-})^{2}  + \frac{u_{-}}{K_{-}}(\nabla\phi_{-})^{2} \right] \nonumber\\
& + \frac{2\rho_{0}^{2}}{(2\pi)^2}g_{3} \int dx~ \cos(\sqrt{8}\phi_{-}), \\
&H_{L} = \frac{2\rho_{0}^{2}}{(2\pi)^2}g_{L} \int dx~ \cos(3\sqrt{2}\phi_{+})\sin(\sqrt{2}\theta_{-}), \nonumber
\end{eqnarray}
where $K_{\pm}=\sqrt{v/(v+g_{1} \pm g_{2})}$ are Luttinger liquid interaction parameters, $u_{\pm}=\sqrt{v(v+g_{1} \pm g_{2})}/2$ are renormalized velocities, and we have defined $g_{L}=\hbar\omega_{B}\frac{(2\pi)^{2}}{2\rho_{0}}$. The scaling dimension of the $g_{3}$ operator is $(2-2K_{-})$, and it corresponds to an anti-symmetric charge density wave. The scaling dimension of the $g_{L}$ is $(2-1/(2K_{-})-9K_{+}/2)$. The renormalization equations defining the fate of the two operators are (see for example \cite{Giamarchi})
\begin{eqnarray}\label{rg}
& \frac{dg_{L}}{d\ell}=\left(2-\frac{1}{2K_{-}}-\frac{9K_{+}}{2}\right)g_{L},~~
\frac{dg_{3}}{d\ell}=\left(2-2K_{-}\right)g_{3}, \nonumber\\
& \frac{dK_{+}}{d\ell}=-\frac{K_{+}^{2}g_{L}^{2}}{2u_{+}^{2}},~~
\frac{dK_{-}}{d\ell}=-\frac{K_{-}^{2}g_{3}^{2}}{2u_{-}^{2}}+\frac{g_{L}^{2}}{2u_{+}^{2}}.
\end{eqnarray}
Note that in the presence of Coulomb interactions $K_{-} \approx 1$, while $K_{+} < 1$. Numerical calculations of (\ref{rg}) show that when the Coulomb interaction is absent, $K_{+} = 1$, the $g_{L}$ is always irrelevant. One has to have $K_{+} \ll 1$ and $g_{L}/u_{+} > g_{3}/u_{-}$ for the $g_{L}$ operator to be the only relevant operator. In further discussions we assume that it is the case.

Let us now discuss other possible non-diagonal matrix elements of (\ref{nondiagonal}), which were skipped before. In the case of, for example, $(5,~-1)$ harmonic matchings, the corresponding term is $\propto \sin(3(\phi_{j}+\phi_{j+1}) + 2(\phi_{j}-\phi_{j+1}) + (\theta_{j}-\theta_{j-1}) )$. In terms of definitions of expression (\ref{two}) and (\ref{rg}), its scaling dimension is $(1/(2K_{-}) + 9K_{+}/2 + 2K_{-})$. This term is always irrelevant because of $K_{-}$, or, in other words, both $\phi_{-}$ and $\theta_{-}$ fields can not order simultaneously. From the analysis above we can conclude that the only relevant operators derived from (\ref{nondiagonal}) are those \textit{with a match of the opposite in sign and equal in magnitude harmonics} (for example $(3,~-3)$ in the Fig. \ref{fig1}).

When the $g_{L}$ dominates, all neighboring momentum states strongly couple via $(3,~-3)$, and the bulk gets gapped. In this case, $\phi_{+}$ field is a two fold degenerate. The kink connecting two ground states carries a charge of $e/3$ \cite{Schulz}. There are only two remaining decoupled harmonics which are located at the edges \cite{Kane,Kane2}. The edges are gapless and chiral, meaning that there is only a right-mover on the right edge, and the left-mover on the left edge \cite{Wen}. Their directions change with the change of the magnetic field direction. The cosine term given by the expression (\ref{laughlin2}) sets the periodicity condition on the edge bosonic field. Namely, the fermion operator describing the edge should be invariant under $\phi_{R(L)} \rightarrow \phi_{R(L)} + 2\pi/3$ transformation, where $\phi_{R(L)}$ is a bosonic field describing the right (left) edge of the system. That selects the choice of the edge fermion operator $\Psi_{R}^{\dag} \propto e^{i3\phi_{R}}, ~~\Psi_{L}^{\dag} \propto e^{-i3\phi_{L}}$. Therefore, the edges carry a fractional charge of $e/3$. Obtained chiral edge states are consistent with Wen's edge description \cite{Wen}.

For the case of $\nu=2/3$ filling factor, one has to consider the same harmonics as in (\ref{laughlin}), but with $m=2$. And the same analysis of relevancy of operators applies. From the first glance, there will be two states on every edge each carrying a charge of $e/3$ so that the overall charge on the edge is $2e/3$. When the filling factor is $\nu=1/2$, the expression for non-diagonal matrix element (\ref{laughlin}) has to be taken as $\int dx~(\Psi_{b,j}^{(n=-1)})^{\dag} \Psi_{b,j+1}^{(n=1)}$. It is a coupling of $(3,~-1)$ harmonics, and from the discussion above it is clear that this fractional state is always forbidden. The symmetric sector, $(+)$ in the definitions of the expression (\ref{two}), is gapless, corresponding to a Fermi liquid. This is consistent with previous theories (for example \cite{half}). From the above discussions we can conclude that the allowed fractional states are
\begin{equation}\label{hierarchy}
\nu = \frac{m}{2n+1},
\end{equation}
where $m$ and $n$ are integers, which is consistent with \cite{Haldane83,Halperin84,Jain89,Read90}.

{\it Conclusions. -} To conclude, this paper proposes an approach to understand the fractional quantum Hall effect. The approach is based on the phenomenological one-dimensional bosonization generalized for the case of two-dimensional electron gas in a strong quantizing magnetic field. The constructed bosonized fermion operator (\ref{bosonizedwf}) of a electron state with a given Landau gauge momentum is represented by its harmonics of the density fluctuation, and covers the entire $x-y$ plane. This fermion operator was used to understand the problem of the FQHE. As an example, it was shown that the $\nu=1/3$ Laughlin state is a mixing of high harmonics of the two fermion operators (\ref{bosonizedwf}) with neighboring momentum. It results in a sine-Gordon model, which gaps the bulk leading to the chiral gapless edges. The overall predicted hierarchy of the fractional charges is consistent with experiments and existing theories \cite{Haldane83, Halperin84, Jain89, Read90}. It is possible to draw an intuitive connection from the presented in this paper formalism to Laughlin wave function and to composite fermions description of the FQHE. We would like to point out, that presented in this paper approach generalizes previous attempts to understand the FQHE within the sliding Luttinger liquid \cite{Kane,Kane2}.

It would be interesting to realize an exact mapping from Laughlin many-body electron wave function to one-dimensional electron stripes, which are the building blocks of the fermion operator (\ref{bosonizedwf}). A more detailed description of the $2/3$ fractional state within the presented formalism is a subject of future research. Also, the development of the non-Abelian fractional states, such as $5/2$ states, is in order.

{\it Acknowledgments. -} I would like to thank my research adviser G.A. Fiete for raising my interest in the problem and for valuable discussions with him. I am thankful to V. Chua, S. Ganeshan, M.M. Glazov, A.A. Greshnov, P. Lecheminant, A.H. MacDonald, A. Slepko, and A.Yu. Zyuzin for helpful discussions. This work is supported by ARO grant W911NF-09-1-0527 and NSF grant DMR-0955778.

\end{document}